\documentstyle[aps,12pt]{revtex}
\begin{document}
\title{Cosmic Strings in Supergravity}
\author{J.R. Morris}
\address{{\it Physics Dept., Indiana University Northwest,}\\
{\it 3400 Broadway, Gary, Indiana 46408}\\
\smallskip\ }
\maketitle

\begin{abstract}
It is pointed out that various types of cosmic string solutions that exist
in nonsupersymmetric and globally supersymmetric theories, such as D-type
gauge strings, F-type global and gauge strings, and superconducting Witten
strings, also exist in supergravity models. When the D term and
superpotential satisfy some simple conditions allowing the determination of
a set of vacuum states with nontrivial topology, the existence of a string
embedded within a supersymmetric vacuum with vanishing cosmological constant
can be inferred. Supergravity also admits other string solutions, some of
which have no counterparts in globally supersymmetric theories.

\smallskip\ 


\bigskip\ 
\end{abstract}

\section{Introduction}

It is now generally recognized that the early Universe may have undergone
symmetry breaking phase transitions which spawned the production of
topological defects\cite{vsbook,ktbook}. Cosmic strings, in particular, may
have played an important role in the subsequent evolution of the universe%
\cite{vsbook,vilenkin}. Furthermore, some realization of supersymmetry,
perhaps in the form of supergravity, could present itself as an effective
theory for the epochs of cosmic string formation. It is therefore natural to
investigate cosmic string solutions in the context of supergravity. Studies
of supergravity domain walls\cite{cvetic1,cvetic4} have indicated that the
gravitational effects in supergravity can play an important role in the
existence and structure of these defects, and we can expect this to be true
for cosmic strings, as well.

The bosonic sector of a theory with global supersymmetry (SUSY) can have a
more complicated form than that of a nonsupersymmetric theory\cite{wbbook},
and this can lead to differences between string solutions in the two
different theories. For example, supersymmetric cosmic string models may
require the participation of more fields than the nonsupersymmetric
counterpart\cite{morriscs}, so that the supersymmetry strings have some
features that are lacking in the nonsupersymmetric versions. Also, string
solutions can emerge from globally supersymmetric theories from a
spontaneous symmetry breaking due to either a $D$ term (D-type strings) or
an $F$ term (F-type strings) in the scalar potential\cite{davistrod,morriscs}%
. Since the scalar potential of a supergravity theory can be more
complicated and may take quite a different appearance than that of a
globally supersymmetric theory with the same form of superpotential, it may
not be immediately obvious whether a given superpotential will generate
scalar potentials in global supersymmetry and supergravity models that will
possess the same types of vacuum manifolds that have the same topology.
Therefore, attention is focused here on the scalar potential and vacuum
manifold resulting from a given superpotential in supergravity, and
comparisons can then be made to the corresponding vacuum manifold arising
from the same superpotential in a globally supersymmetric theory containing
the same matter chiral superfields. Under certain simple conditions for the
D term potential and the superpotential, the same type of string solution
will be admitted by both the globally supersymmetric and supergravity
theories. However, if these conditions on the superpotential are not met,
the two theories can yield quite different solutions. An example can be seen
for the case of a constant superpotential, which is dynamically irrelevant
in global SUSY, but can lead to spontaneous symmetry breaking in
supergravity. Since supergravity can accommodate a negative cosmological
constant and global SUSY can not, there are some string solutions admitted
by supergravity that have no counterparts in global SUSY. This result can be
viewed as arising from gravitational effects that are included in
supergravity, but not in global SUSY.

We concern ourselves mostly with the situation where the cosmological
constant $\Lambda $ vanishes. For the case of a global SUSY theory with a
superpotential $W$, we conclude that a topological string solution is
admitted if the vacuum manifold, characterized by the field vacuum
configuration $\varphi =\{\varphi ^i\}$, possesses a nontrivial topology
(such as the topology of $S^1$) and can be obtained from the conditions (i) $%
\left\langle D\right\rangle =0$ and (ii) $\left\langle \partial W/\partial
\phi ^i\right\rangle =0$. For the case of supergravity, a string solution is
admitted if the same kind of vacuum configuration can be obtained from the
conditions (i) $\left\langle D\right\rangle =0$, (ii) $\left\langle \partial
W/\partial \phi ^i\right\rangle =0$, and (iii) $\left\langle W\right\rangle
=0$. These conditions are seen to be necessary and sufficient to locate $%
\Lambda =0$ vacuum states with unbroken supersymmetry. They can be quite
useful since the functions $D$ and $W$ are much easier to examine than is
the scalar potential.

In the next section we focus upon the bosonic sector and scalar potential of
global SUSY theories, and recall some of the various types of string
solutions that occur there. Specifically, we present examples of the D-type
gauge string, the F-type global or gauge string, and the superconducting
Witten string\cite{witten}, each of which is surrounded by supersymmetric
vacuum with zero cosmological constant, $\Lambda =0$. The scalar potential
of supergravity is presented in section III, and we show that the same types
of $\Lambda =0$ string solutions exist here as well, along with $\Lambda
\neq 0$ strings. The minimal K\"{a}hler potential is extended to a general
form in section IV, and a brief summary forms section V.

\section{Global Supersymmetry Strings}

Let us consider a globally supersymmetric theory with interacting chiral
superfields $\Phi ^i$ and a $U(1)$ vector superfield $A$. The matter chiral
supermultiplets can be represented as $\Phi ^i=(\phi ^i,\psi ^i,F^i)$ and
the vector multiplet by $A=(A^\mu ,\lambda _\alpha ,\bar{\lambda}_{\dot{%
\alpha}},D)$, where $F^i$ and $D$ are the bosonic auxiliary fields, $\psi ^i$
is the matter fermion, and $\lambda _{\alpha \text{ }}$ is the photino.%
\footnote{%
Aside from a spacetime metric with signature given by $(+,-,-,-)$, we use,
for the most part, the notation and conventions of Wess and Bagger\cite
{wbbook}. Units are chosen where $M=M_P/\sqrt{8\pi }$ $=1$, where $%
M_P=G^{-1/2}$ is the Planck mass. Factors of $M$ can be reintroduced by
dimensional considerations.} For the present we choose the minimal
K\"{a}hler potential $K=\bar{\phi}^i\phi ^i=\sum\limits_i\phi ^{*i}\phi ^i$
which yields the diagonal K\"{a}hler metric $K_{i\bar{j}}=\frac{\partial ^2K%
}{\partial \phi ^i\partial \bar{\phi}^j}=\delta _{ij}=K^{i\bar{j}}$ and
canonical kinetic terms. The bosonic sector of the theory is described by
the Lagrangian 
\begin{equation}
L_B=K^{i\bar{j}}(D_\mu \phi ^i)(D^\mu \phi ^j)^{*}-\frac 14(F_{\mu \nu })^2-V
\label{e1}
\end{equation}

\noindent with a sum over $i$ implied, and where $D_\mu \phi ^i=(\partial
_\mu +igQ_iA_\mu )\phi ^i$ is the gauge covariant derivative of the field $%
\phi ^i$ with a $U(1)$ charge $Q_i$, and $F_{\mu \nu }=\partial _\mu A_\nu
-\partial _\nu A_\mu $. The scalar potential is 
\begin{equation}
V=\sum\limits_i\left| F^i\right| ^2+\frac 12D^2=\sum\limits_i\left| \frac{%
\partial W}{\partial \phi ^i}\right| ^2+\frac 12\left[ \xi +g\sum\limits_iQ_i%
\bar{\phi}^i\phi ^i\right] ^2,  \label{e3}
\end{equation}
\noindent where the holomorphic function $W=W(\phi ^i)$ is the
superpotential and the constant $\xi $ comes from a Fayet-Iliopoulos term%
\cite{fayet2} that has been included and $\bar{F}^i=-\frac{\partial W}{%
\partial \phi ^i}$, $D=-\left[ \xi +g\sum\limits_iQ_i\bar{\phi}^i\phi
^i\right] $. A superpotential of the form $W=W_0+a_i\phi ^i+b_{ij}\phi
^i\phi ^j+c_{ijk}\phi ^i\phi ^j\phi ^k$ allows for renormalizability, and
since the constant $W_0$ is dynamically irrelevant, we can set it equal to
zero.

Let us use the derivative notation $X_i=\partial X/\partial \phi ^i$, $X_{%
\bar{j}}=\partial X/\partial \bar{\phi}^j$, $\bar{X}_{\bar{j}}=\partial \bar{%
X}/\partial \bar{\phi}^j$, $X_{i\bar{j}}=\partial ^2X/\partial \phi
^i\partial \bar{\phi}^j$, $X_iX_{\bar{i}}=\sum\limits_iX_iX_{\bar{i}}$ ,
etc. for some function $X(\phi ,\bar{\phi})$, with a sum over repeated
indices unless otherwise stated. The vacuum expectation value (vev) $%
\left\langle \phi ^i\right\rangle =\varphi ^i$ is located at the minimum of $%
V$ where 
\begin{equation}
V_i=\bar{W}_{\bar{k}}W_{ki}+DD_i=\bar{F}^kF_i^k+DD_i  \label{e4}
\end{equation}

\noindent vanishes. We also note that $V\geq 0$ so that a negative
cosmological constant does not appear. The vacuum state is supersymmetric if 
$V(\varphi )=0$, but supersymmetry is spontaneously broken by the vacuum if $%
V(\varphi )>0$. From (\ref{e4}) it is seen that a nonzero vev $\varphi
^i\neq 0$ can develop from either the $F$ term or the $D$ term in $V$,
resulting in either F-type or D-type strings\cite{davistrod}. Abelian F-type
and D-type strings in global SUSY theories are described and discussed in
ref.\cite{davistrod}, to which the reader is referred, and a global SUSY
model of a local superconducting Witten string, which is an F-type string,
is discussed in ref.\cite{morriscs}. Below we list some of the basic
features of these string solutions, and in the next section these types of
strings will be re-examined in the context of supergravity.

\smallskip\ 

{\it D-type string:} An example of a D-type string is given in\cite
{davistrod}. The model contains a primary charged chiral superfield $\Phi $
that is involved with the spontaneous symmetry breaking, along with other
charged chiral superfields (to avoid a gauge anomaly), and possibly neutral
chiral superfields. The various chiral superfields interact in such a way
that only the one complex scalar field $\phi $ with $U(1)$ charge $Q$
develops an expectation value, with $W=0$ and $W_i=0$ in the vacuum state.
By including a Fayet-Iliopoulos term, setting $\xi =-gQ\eta ^2$, and setting
to zero the charged scalar fields with vanishing vacuum values, the scalar
potential due to the symmetry breaking field $\phi $ that arises from the $D$
term is $V_D=\frac 12g^2Q^2(\bar{\phi}\phi -\eta ^2)^2$. This is the type of
potential found in the ordinary broken symmetric Abelian-Higgs model, which
admits a Nielsen-Olesen cosmic $U(1)$ gauge string\cite{nielsen}. In the
vacuum state, outside the string, $|\phi |=$ $|\varphi |=\eta $ and $%
D(\varphi )=0$ with $V(\varphi )=0$, which implies that supersymmetry is not
broken in the vacuum, and the cosmological constant vanishes, $\Lambda =0$.
The structure of the vacuum manifold is determined entirely by the $D$ term
in the scalar potential.

\smallskip\ 

{\it F-type string:} An F-type string model can be built from one neutral
superfield ${\cal Z}=(Z,\psi _Z,F_Z)$ and two charged chiral superfields $%
\Phi _{\pm }=(\phi _{\pm },\psi _{\pm },F_{\pm })$ with $Q_{+}+Q_{-}=0=Q_Z$.
A Fayet-Iliopoulos term is not included and the superpotential is taken tobe 
$W=\lambda {\cal Z}(\Phi _{+}\Phi _{-}-\eta ^2)$. The lowest energy state of
the theory that spontaneously breaks the $U(1)$ symmetry is a supersymmetric
vacuum state with $V=0$, determined by the conditions $F^i=0$, $D=0$, and is
characterized by $|\left\langle \phi _{\pm }\right\rangle |=|\varphi _{\pm
}|=\eta $, $\left\langle Z\right\rangle =0$. Using the simplifying ansatz $%
\phi _{+}=\phi $, $\phi _{-}=\bar{\phi}$, the $D$ term vanishes and scalar
potential reduces to 
\begin{equation}
V=\lambda ^2(\bar{\phi}\phi -\eta ^2)^2+2\lambda ^2\bar{Z}Z(\bar{\phi}\phi ).
\label{e7}
\end{equation}

\noindent For the case in which $Z$ is set equal to its vacuum value of
zero, the model reduces to a broken symmetric Abelian-Higgs model, admitting
a $U(1)$ gauge string associated with the field $\phi $. Since $\phi =\phi
_{+}=\bar{\phi}_{-}$, we see that the two fields $\phi _{\pm }$ conspire to
form this F-type string. In this case the vacuum manifold for $\phi $ and $Z$
is determined entirely by the $F$ terms, where $F_{+}=\lambda Z\phi
_{-}\rightarrow \lambda Z\bar{\phi}$, $F_{-}=\lambda Z\phi _{+}\rightarrow
\lambda Z\phi $, $F_Z=\lambda (\phi _{+}\phi _{-}-\eta ^2)\rightarrow
\lambda (\bar{\phi}\phi -\eta ^2)$. Again, the vacuum is supersymmetric and $%
\Lambda =0$. (Note that we could replace the $U(1)$ gauge symmetry in this
model with a global $U(1)$ symmetry, remove the vector supermultiplet, and
take $g\rightarrow 0$ to get a model of a global string.)

\smallskip\ 

{\it Superconducting Witten string:} We can construct a model of a $%
U(1)\times U(1)^{\prime }$ superconducting global string, with long range
gauge fields outside of the string, from two complex scalar fields $\sigma
_{\pm }$ transforming nontrivially under a global group $U(1)^{\prime }$,
two complex scalars $\phi _{\pm }$ transforming nontrivially under a local
group $U(1)$, and a neutral scalar $Z$. The fields $\sigma _{\pm }$ have $%
U(1)^{\prime }$ global charges $Q_{\pm }$ with $Q_{+}+Q_{-}=0$, and the
fields $\phi _{\pm }$ have $U(1)$ local charges $q_{\pm }$ with $%
q_{+}+q_{-}=0$. A Fayet-Iliopoulos term is not included and the
superpotential, in terms of the scalar fields, is taken to be 
\begin{equation}
W=\lambda Z(\sigma _{+}\sigma _{-}-\eta ^2)+(cZ+m)\phi _{+}\phi _{-}.
\label{e7a}
\end{equation}

\noindent The scalar potential is $V=\sum\limits_k|W_k|^2+\frac 12D^2$,
where $\,\,D=gq_{+}(\bar{\phi}_{+}\phi _{+}-\bar{\phi}_{-}\phi _{-})$, and
the individual $W_k$ terms are given by 
\begin{equation}
\begin{array}{ll}
\frac{\partial W}{\partial \sigma _{\pm }} & =\lambda Z\sigma _{\mp } \\ 
\frac{\partial W}{\partial \phi _{\pm }} & =(cZ+m)\phi _{\mp } \\ 
\frac{\partial W}{\partial Z} & =\lambda (\sigma _{+}\sigma _{-}-\eta
^2)+c\phi _{+}\phi _{-}.
\end{array}
\label{e10}
\end{equation}

The supersymmetric vacuum state with $V=0$ that spontaneously breaks the $%
U(1)^{\prime }$ symmetry but respects the $U(1)$ symmetry is located where $%
W_k=0$ and $D=0$, i.e. where the fields take values $\sigma _{+}\sigma
_{-}=\eta ^2$, $\phi _{+}=\phi _{-}=0$, and $Z=0$. One can adopt the ansatz $%
\sigma _{+}=\sigma $, $\sigma _{-}=\bar{\sigma}$, $\phi _{+}=\phi $, $\phi
_{-}=\bar{\phi}$, so that the scalar potential can be written in terms of
the fields $\sigma $, $\phi $, and $Z$, and a superconducting string
solution can be inferred. Here again, $\Lambda =0$, since the vacuum is
globally supersymmetric.

A point to be made here, which will be useful in the next section, is that
for each of the examples above for $\Lambda =0$ strings, we have not only $%
W_k(\varphi )=0$, but also $W(\varphi )=0$.

\section{Supergravity Strings}

Now consider a supergravity theory accommodating matter chiral superfields $%
\Phi ^i=(\phi ^i,\psi ^i,F^i)$ and the vector multiplet $A=(A^\mu ,\lambda
_\alpha ,\bar{\lambda}_{\dot{\alpha}},D)$. We again choose the minimal
K\"{a}hler potential $K=\bar{\phi}^i\phi ^i$ so that $K_i=\bar{\phi}^i$, $K_{%
\bar{i}}=\phi ^i$, and $K_{i\bar{j}}=K^{i\bar{j}}=\delta _{ij}$. The bosonic
sector of the theory is described by 
\begin{equation}
e^{-1}{\cal L}_B=-\frac 12R+K_{i\bar{j}}(D_\mu \phi ^i)(D^\mu \phi ^j)^{*}-%
\frac 14F^{\mu \nu }F_{\mu \nu }-{\cal V},  \label{e13}
\end{equation}

\noindent where $e=(-\det g_{\mu \nu })^{1/2}$. Upon defining $D_iW=W_i+K_iW$%
, the scalar potential can be written as 
\begin{equation}
{\cal V}=e^K{\cal U}+\frac 12D^2,  \label{e14}
\end{equation}

\noindent where 
\begin{equation}
{\cal U}=K^{i\bar{j}}(D_iW)(D_jW)^{*}-3W^{*}W=\left| D_iW\right| ^2-3\left|
W\right| ^2  \label{e15}
\end{equation}

\noindent with an implied sum over the repeated indices $i$ and $j$, and $%
D=g\sum\limits_iQ_i\bar{\phi}^i\phi ^i$, as before. For the K\"{a}hler
potential $K=\bar{\phi}^i\phi ^i$ we have the operator $D_i=\partial
/\partial \phi ^i+\bar{\phi}^i$, so that we can also write ${\cal U}$ in
terms of $W$ and $W_i$ as 
\begin{equation}
{\cal U}=\left| W_i\right| ^2+\left( \left| \phi ^i\right| ^2-3\right)
\left| W\right| ^2+(\phi ^iW_i\bar{W}+\bar{\phi}^i\bar{W}_iW).  \label{e16}
\end{equation}

We notice that the supergravity scalar potential, having gravitational
contributions, is more complicated than the scalar potential for global
SUSY, and that ${\cal V}$ can be positive, negative, or zero. Whereas the
signal of spontaneous supersymmetry breaking is given by $W_i\neq 0$ or $%
D\neq 0$ in global SUSY, in supergravity the signal is given by $D_iW\neq 0$
or $D\neq 0$.

If we reinsert factors of $M=M_P/\sqrt{8\pi }$ by introducing the constant $%
\kappa =M^{-1}$ with $W\rightarrow \kappa ^3W$, $D_iW\rightarrow \kappa
^2W_i+\kappa ^4\bar{\phi}^iW$, and ${\cal V}\rightarrow \kappa ^{-4}{\cal V}$%
, we can write the scalar potential as 
\begin{equation}
{\cal V}=\exp (\kappa ^2\bar{\phi}^i\phi ^i)\left\{ \left| W_i+\kappa ^2\bar{%
\phi}^iW\right| ^2-3\kappa ^2\left| W\right| ^2\right\} +\frac 12D^2.
\label{e17}
\end{equation}

\noindent In the small $\kappa $ (large $M$)\thinspace limit we could expand 
${\cal V}$ as ${\cal V}\approx \left| W_i\right| ^2+\frac 12D^2$ with $%
O(\kappa ^2)$ corrections, i.e. the supergravity scalar potential should
just be the global SUSY potential with $O(\kappa ^2)$ corrections, which
would presumably be small. This would lead us to believe that, for low
symmetry breaking energy scales, any superpotential that gives rise to
string solutions in global SUSY will give rise to the same string solutions
in supergravity, and that if no strings are predicted by global SUSY, then
none will be predicted by supergravity either. This seems reasonable if all
field vacuum values are small compared to $M$, but this need not be the
case, in general. Also, a more general K\"{a}hler potential could be used,
to which global SUSY and supergravity can have different sensitivities. This
situation is examined later, where the minimal K\"{a}hler potential is
replaced by a general K\"{a}hler potential. However, we can use the {\it %
exact} form of ${\cal V}$ to see that under certain simple conditions
involving the D term potential and the superpotential, a $\Lambda =0$ global
SUSY string with unbroken supersymmetry in the vacuum will also show up in
supergravity as a $\Lambda =0$ supergravity string with unbroken
supersymmetry in the vacuum, {\it regardless} of the scale of the symmetry
breaking. This is not guaranteed to be the case, however, if these
conditions are not met.

\subsection{Vacuum states}

The vacuum states, labelled by $\left\langle \phi ^i\right\rangle =\varphi
^i $, are located where ${\cal V}_i=\partial {\cal V}/\partial \phi ^i$
vanishes. From (\ref{e14}) 
\begin{equation}
{\cal V}_i=e^K({\cal U}_i+\bar{\phi}^i{\cal U})+D\partial D/\partial \phi ^i.
\label{e18}
\end{equation}

\noindent From (\ref{e16}) we can display ${\cal U}_i$ as 
\begin{equation}
{\cal U}_i=(\bar{W}_{\bar{k}}+\phi ^k\bar{W})W_{ki}+\left[ \left( \left|
\phi ^k\right| ^2-2\right) \bar{W}+\bar{\phi}^k\bar{W}_{\bar{k}}\right] W_i+%
\bar{\phi}^i\left| W\right| ^2.  \label{e19}
\end{equation}

\noindent Solutions to ${\cal V}_i=0$ are given by 
\begin{equation}
{\cal U}_i+\bar{\phi}^i{\cal U}=0,\,\,\,\,\,\,\,\,\,\,D\frac{\partial D}{%
\partial \phi ^i}=(gQ_i\bar{\phi}^i)D=0.  \label{e20}
\end{equation}

\noindent We note that the field $\phi ^i$ develops a nonzero vev if the
curvature of ${\cal V}$ in the $\phi ^i$ direction (i.e. the mass term for $%
\phi ^i$) is negative at the origin, e.g. ${\cal V}_{i\bar{i}%
}^{(0)}=(\partial ^2{\cal V}/\partial \phi ^i\partial \bar{\phi}^i)|_{\phi
=0}<0$, where $\phi $ collectively represents the set $\{\phi ^k\}$.
Assuming a superpotential of the form 
\begin{equation}
W=W_0+a_i\phi ^i+b_{ij}\phi ^i\phi ^j+c_{ijk}\phi ^i\phi ^j\phi ^k+\cdot
\cdot \cdot  \label{e21}
\end{equation}

\noindent we can calculate 
\begin{eqnarray}
{\cal V}_{i\bar{j}}^{(0)} &=&\delta _{ij}{\cal U}^{(0)}+{\cal U}_{i\bar{j}%
}^{(0)}+\frac 12\left[ \partial ^2(D^2)/\partial \phi ^i\partial \bar{\phi}%
^j\right] |_0  \label{e21a} \\
&=&\delta _{ij}\left( \bar{a}_ka_k-2\left| W_0\right| ^2\right) -a_i\bar{a}%
_j+4b_{ki}\bar{b}_{kj}+gQ_i\delta _{ij}\xi ,  \nonumber
\end{eqnarray}
\noindent where $\xi $ comes from a Fayet-Iliopoulos term, giving 
\begin{equation}
{\cal V}_{i\bar{i}}^{(0)}=-2\left| W_0\right| ^2-\left| a_i\right|
^2+\sum\limits_k\left( \left| a_k\right| ^2+4\left| b_{ki}\right| ^2\right)
+gQ_i\xi .  \label{e22}
\end{equation}

\noindent The corresponding curvature for the case of global SUSY is 
\begin{equation}
V_{i\bar{i}}^{(0)}=4\sum\limits_k\left| b_{ki}\right| ^2+gQ_i\xi .
\label{e22a}
\end{equation}
We can therefore see that whereas the constant term $W_0$ is innocuous in
global SUSY, in a supergravity model it contributes toward a destabilization
of the normal vacuum state $\left\langle \phi ^i\right\rangle =0$, so that
for a large enough value of $\left| W_0\right| $, spontaneous symmetry
breaking is induced. Therefore, although a particular superpotential may not
lead to a symmetry breaking in global SUSY, due to an invisibility of $W_0$,
the same superpotential can lead to a symmetry breaking in supergravity.
However, we will see that when the cosmological constant vanishes and the D
term and superpotential satisfy certain simple conditions, the vacuum states
in global SUSY will coincide with those in supergravity.

\subsection{$\Lambda =0$ strings}

From (\ref{e14}), we can note that for a spontaneously broken symmetry the
cosmological constant will vanish, i.e. $\left\langle {\cal V}\right\rangle
=\Lambda =0$, provided that $\left\langle {\cal U}\right\rangle
=\left\langle D\right\rangle =0$. In this case, by (\ref{e20}), the
condition ${\cal V}_i=0$ is satisfied when the conditions ${\cal U}_i=0$ and 
$(gQ_i\bar{\phi}^i)D=0$ are simultaneously satisfied. For $\left\langle \phi
^i\right\rangle \neq 0$, $\left\langle D\right\rangle $ can vanish either
because of symmetry or because of a symmetry breaking Fayet-Iliopoulos term.
Assuming this to be the case, we are left with the condition ${\cal U}_i=0$.
This condition will be met automatically if the field expectation values $%
\left\langle \phi ^i\right\rangle =\varphi ^i$ can be obtained from the
conditions $W=0$ and $W_i=0$ when $\{\phi ^k\}=\{\varphi ^k\}$. Therefore,
if a globally supersymmetric theory possesses a nontrivial vacuum
configuration $\varphi =\{\varphi ^i\}$ that can be extracted from the
conditions 
\begin{equation}
(i)\,\,\left\langle D\right\rangle =0,\,\,\,\,\,\,\,(ii)\,\,\left\langle
W_i\right\rangle =W_i(\varphi )=0,\,\,\,\,\,\,\,(iii)\,\,\left\langle
W\right\rangle =W(\varphi )=0,  \label{e23}
\end{equation}

\noindent and as a result admits a string solution, then the same vacuum
configuration $\varphi $ appears in the supergravity theory and hence a
string solution appears there, although, due to differences in the forms of
the scalar potentials and fermionic couplings, the two string solutions will
in general be quantitatively different. We can briefly look at the D-type
string, the F-type string, and the superconducting Witten string of the
previous section for specific examples.

\subsubsection{D-type string}

As in the global SUSY case, the superpotential for this model satisfies $%
\left\langle W\right\rangle =0$, $\left\langle W_i\right\rangle =0$. A
Fayet-Iliopoulos term is included, and the scalar field $\phi $ that is
nonvanishing in the vacuum gives a nontrivial contribution to the D-term, so
that we can write, as in the case of global SUSY, (where we again set to
zero the scalar fields with vanishing vacuum values)  $D=gQ\bar{\phi}\phi
+\xi =gQ(\bar{\phi}\phi -\eta ^2)$. For the vacuum state where $\left|
\left\langle \phi \right\rangle \right| =\left| \varphi \right| =\eta $, we
have, as before,  $\left\langle D\right\rangle =0$, $\left\langle
W\right\rangle =0$, $\left\langle W_i\right\rangle =0$, indicating the
existence of a D-type string in the supergravity theory. Also, since $D_iW=0$
in the vacuum state, supersymmetry is respected in the vacuum. Furthermore,
from (\ref{e22}), we see that ${\cal V}_{\phi \bar{\phi}}^{(0)}=-(gQ\eta )^2$%
.

\subsubsection{F-type string}

For this model we have $D=gQ_{+}(\bar{\phi}_{+}\phi _{+}-\bar{\phi}_{-}\phi
_{-})$ and the superpotential is $W=\lambda Z(\phi _{+}\phi _{-}-\eta ^2)$.
The $W_i$ are given by $W_Z=\lambda (\phi _{+}\phi _{-}-\eta ^2)$, $W_{\pm
}=\lambda Z\phi _{\mp }$, so that $W_i=0$ is satisfied by $Z=0$, $\phi
_{+}\phi _{-}=\eta ^2$, and consequently $W=0$. The condition $\left\langle
D\right\rangle =0$ implies that $\left| \left\langle \phi _{+}\right\rangle
\right| =\left| \left\langle \phi _{-}\right\rangle \right| \equiv \left|
\left\langle \phi \right\rangle \right| $, in accordance with the ansatz
implemented previously. Since $D_iW|_{\phi =\varphi }=W_i(\varphi )+\varphi
^iW(\varphi )=0$, we have an F-type string surrounded by supersymmetric
vacuum.

\subsubsection{Superconducting Witten string}

The superpotential is given by (\ref{e7a}), and by (\ref{e10}) we have $%
\sigma _{+}\sigma _{-}=\eta ^2$, $\phi _{\pm }=0$, $Z=0$ as solutions of $%
W_i=0$, and consequently $W(\varphi )=0$. Requiring $\left\langle
D\right\rangle =0$ gives $\left| \left\langle \phi _{+}\right\rangle \right|
=\left| \left\langle \phi _{-}\right\rangle \right| $ as in our ansatz. $%
D_iW=0$ in the vacuum, so that the supergravity theory admits a solution
describing a superconducting string embedded in supersymmetric vacuum.

\subsection{Anti-de Sitter ($\Lambda <0$) strings}

Global SUSY can not accommodate a negative cosmological constant (anti-de
Sitter spacetime), since $V=\left| W_k\right| ^2+\frac 12D^2\geq 0$, but
supergravity can, and consequently supergravity can admit $\Lambda <0$
string solutions describing strings embedded in anti-de Sitter spacetime
that have no counterparts in global SUSY. As a simple example we take a
model with a single chiral superfield $\Phi $ invariant under a global $U(1)$
symmetry, so that $D=0$. We choose a real constant superpotential $W=W_0$.
By (\ref{e14}) and (\ref{e16}) the scalar potential takes the form 
\begin{equation}
{\cal V}=W_0^2e^K\left( \left| \phi \right| ^2-3\right) .  \label{e24}
\end{equation}

\noindent The vacuum state is located by $\left| \phi \right| =\sqrt{2}$ .
The global $U(1)$ symmetry is broken, and a global string solution\cite
{vilev} is admitted. Since $D_\phi W=\bar{\phi}W_0$ does not vanish in the
vacuum, supersymmetry is also broken. However, in the core of the string
where $\phi \rightarrow 0$, supersymmetry is apparently restored. This seems
to be the opposite of the case with the $\Lambda =0$ strings above, where
the vacuum is supersymmetric and supersymmetry is broken in the string core.
For this supermassive anti-de Sitter string, $\left| \left\langle \phi
\right\rangle \right| \sim M$, and gravity is expected to play an important
role.

\section{General K\"{a}hler Potential and $\Lambda =0$ Strings}

For a general K\"{a}hler potential $K(\phi ,\bar{\phi})$ the scalar
potential for global SUSY is 
\begin{equation}
V=K^{i\bar{j}}W_i\bar{W}_{\bar{j}}+\frac 12D^2,  \label{e25}
\end{equation}

\noindent where $K^{i\bar{j}}=(K^{-1})_{i\bar{j}}$ and $D=(g\sum%
\limits_iQ_iK_i\phi ^i+\xi )$. This potential is minimized by a set of field
configurations $\varphi =\{\varphi ^i\}$ when 
\begin{equation}
V_k=(K^{i\bar{j}})_kW_i\bar{W}_{\bar{j}}+K^{i\bar{j}}W_{ik}\bar{W}_{\bar{j}%
}+D\frac{\partial D}{\partial \phi ^k}  \label{e25a}
\end{equation}

\noindent vanishes. The scalar potential for supergravity is 
\begin{equation}
{\cal V}=e^K{\cal U+}\frac 12D^2=e^K\left[ K^{i\bar{j}}(D_iW)(D_jW)^{*}-3%
\left| W\right| ^2\right] +\frac 12D^2,  \label{e26}
\end{equation}

\noindent where $D_iW=W_i+K_iW$. This scalar potential is minimized when 
\begin{equation}
{\cal V}_k=e^K({\cal U}_k+K_k{\cal U})+D\frac{\partial D}{\partial \phi ^k}
\label{e26a}
\end{equation}

\noindent vanishes, with 
\begin{equation}
\begin{array}{ll}
{\cal U}_k= & (\partial _kK^{i\bar{j}})(D_iW)(D_jW)^{*} \\ 
& +K^{i\bar{j}}\left[ (\partial _kD_iW)(D_jW)^{*}+(D_iW)\partial
_k(D_jW)^{*}\right] -3\bar{W}W_k,
\end{array}
\label{e27}
\end{equation}

\noindent where $\partial _k=\partial /\partial \phi ^k$.

Now let us assume that $D$ vanishes in the vacuum, $\left\langle
D\right\rangle =0$, and that the cosmological constant vanishes, $\Lambda =0$%
. For global SUSY, $\Lambda =0$ requires that $W_i$ vanish in the vacuum,
i.e. the vacuum values $\left\langle \phi ^i\right\rangle =\varphi ^i$ can
be determined from the condition $W_i(\varphi )=0$. If we furthermore
require that $W(\varphi )=0$, then we see that $(D_iW)|_\varphi =0$, and
therefore ${\cal U}=0$ and ${\cal U}_k=0$, so that ${\cal V}$ is minimized
by the vacuum field configuration $\varphi $. We therefore conclude that if
the superpotential $W$ possesses a global or local $U(1)$ symmetry which is
spontaneously broken by the vacuum, and (i) $\left\langle D\right\rangle =0$
and (ii) $W_i(\varphi )=0$, then a $\Lambda =0$ string solution exists in
the global SUSY theory. If, in addition, we have that (iii) $W(\varphi )=0$,
then a $\Lambda =0$ string solution exists in the supergravity theory.
(However, it does not necessarily follow that a $\Lambda =0$ supergravity
string solution does {\it not} exist if these conditions are {\it not}
satisfied.) We note that although these conditions are stated for a general
K\"{a}hler potential, the condition $\left\langle D\right\rangle =0$ may
constrain the form of $K$ when a local $U(1)$ symmetry is present.

Actually, the conditions (i) - (iii) above are {\it necessary} and {\it %
sufficient} conditions to locate the $\Lambda =0$ vacuum states of the
theory with {\it unbroken supersymmetry}. This can be easily seen by
noticing that unbroken supersymmetry requires that $\left\langle
D\right\rangle =0$ and $\left\langle D_iW\right\rangle =0$. Then, from (\ref
{e15}), it follows that $\left\langle {\cal U}\right\rangle =-3\left\langle
\left| W\right| ^2\right\rangle $, and therefore $\Lambda =0$ implies that $%
\left\langle W\right\rangle =W(\varphi )=0$. Since $D_iW=W_i+K_iW=0$ in the
vacuum, it follows that $\left\langle W_i\right\rangle =0$. If conditions
(i) - (iii) are simultaneously satisfied, then they necessarily locate the $%
\Lambda =0$ supersymmetric vacuum states of a theory. If this set of vacuum
states has a vacuum manifold characterized by a topology where the first
homotopy group is nontrivial, then a topological string solution must exist.
However, if the conditions (ii) or (iii) do not hold simultaneously, then
the vacuum states of the theory do not simultaneously have unbroken
supersymmetry and $\Lambda =0$. String solutions may exist, but they will
not be $\Lambda =0$ strings embedded within supersymmetric vacuum.

Of course, we expect that if the global SUSY or supergravity theory
possesses a $\Lambda =0$ string solution, then the K\"{a}hler transformed
theory will also have a string solution, since the bosonic Lagrangians are
invariant under K\"{a}hler transformations. I.e. for global SUSY $L_B$ and $%
V $ are invariant under the K\"{a}hler transformation\footnote{%
We avoid a K\"{a}hler gauge where $G(\phi ,\bar{\phi})=K(\bar{\phi}\phi
)+\ln P+\ln \bar{P}$, since $G$ becomes undefined when $P=0$.} $K(\phi ,\bar{%
\phi})\rightarrow G(\phi ,\bar{\phi})=K+F(\phi )+\bar{F}(\bar{\phi})$ and $%
W\rightarrow W$, where $F(\phi )$ is an analytic function of the $\phi ^i$.
For supergravity, however, ${\cal L}_B$ and ${\cal V}$ are invariant under
the transformation $K\rightarrow G$ provided that $W\rightarrow \tilde{W}%
=e^{-F}W$. If $\left\langle D\right\rangle $ remains zero under a K\"{a}hler
transformation, and if $W(\varphi )=0$, $W_i(\varphi )=0$, then $\tilde{W}%
_i(\varphi )=0$ and $\tilde{W}(\varphi )=0$, provided that $F_i(\varphi )$
is finite, so that the conditions for $W$ are also satisfied for $\tilde{W}$%
. However, if we consider the K\"{a}hler {\it inequivalent} theories
generated by the replacements $K\rightarrow G$, $W\rightarrow W$, or $%
K\rightarrow K$, $W\rightarrow \tilde{W}$, then we see that the conditions
on the superpotential can still be maintained, since $\tilde{W}(\varphi
)=e^{-F}W(\varphi )=0$ and $\tilde{W}_i(\varphi )=e^{-F}[W_i(\varphi
)-F_iW(\varphi )]=0$. In other words, {\it different} theories that have the 
{\it same} superpotential, but {\it different} K\"{a}hler potentials, or 
{\it different} theories that have the {\it same} K\"{a}hler potential but 
{\it different} superpotentials $W$ and $\tilde{W}$ related by $\tilde{W}%
=e^{-F}W$ can yield the same type of string solution, provided that the
conditions (i) - (iii) above hold. For example, for an F-type global string
model with a superpotential $W=\lambda Z(\phi _{+}\phi _{-}-\eta ^2)$, with $%
\left| \left\langle \phi _{\pm }\right\rangle \right| =\eta $, $\left\langle
Z\right\rangle =0$, conditions (i) -(iii) are satisfied, with $D=0$, so that
any other theory generated by a superpotential $\tilde{W}=\exp [-F(Z,\phi
_{+},\phi _{-})]W$ should also yield a global string solution, assuming that 
$\left\langle F_i\right\rangle $ is finite.. Since $F(\phi )$ is arbitrary,
we see that there are an infinite number of superpotentials that can
generate the same type of string solution. (Renormalizability considerations
in global SUSY, however, can reduce the spectrum of acceptable
superpotentials.)

\section{Summary}

Cosmic strings may have been produced during symmetry breaking phase
transitions in the early universe. It is also possible that supersymmetry,
perhaps in the form of supergravity, was effectively realized at early
epochs. It is then natural to examine cosmic strings in supersymmetric
theories. To determine whether a field theory will yield a topological
string solution, it is necessary to have information about the topology of
the vacuum manifold. This information is embedded within the scalar
potential, and so it is sufficient to focus attention upon this function.
The scalar potential in a supersymmetric theory is constructed from terms
that depend upon the functions $D$, $W$, and $K$, where $W$ is the
superpotential and $K$ is the K\"{a}hler potential. The functions $D$ and $W$
can yield useful information about the vacuum states of the theory if these
functions satisfy certain simple conditions. This can be advantageous, since
the functions $D$ and $W$ are generally much simpler in structure than is
the scalar potential. The existence of cosmic string solutions in a
supersymmetric theory can be inferred when a topologically nontrivial vacuum
manifold can be extracted directly from these functions.

The scalar potentials of global SUSY and supergravity have different forms
for a given superpotential, with the supergravity version being generally
more complicated than the global SUSY one. Therefore, it is not immediately
obvious whether the existence of a string solution in a global SUSY theory
implies the existence of a string solution in the supergravity theory or
vice versa. In fact, because supergravity can support a negative
cosmological constant and global SUSY can not, there can be string solutions
in supergravity that have no counterparts in global SUSY. Also, even in the
case of $\Lambda =0$, if the energy scale of the symmetry breaking is large
(say $\varphi \sim M$), then the gravitational effects inherent in
supergravity may be important.

From an examination of the structure of scalar potentials the following
conclusions have been drawn for the situation where the cosmological
constant $\Lambda $ vanishes. If a vacuum manifold with nontrivial topology
(nontrivial first homotopy group) can be determined from the conditions (i) $%
\left\langle D\right\rangle =0$ and (ii) $\left\langle \partial W/\partial
\phi ^i\right\rangle =0$, then a topological string solution is admitted in
a globally supersymmetric theory. This will be a $\Lambda =0$ string
embedded within supersymmetric vacuum. If, in addition, the condition (iii) $%
\left\langle W\right\rangle =0$ is satisfied, then the existence of a
topological string solution in a supergravity theory can be inferred.
Therefore, when conditions (i) - (iii) are satisfied, a $\Lambda =0$ string
solution (that asymptotically relaxes into a vacuum with unbroken
supersymmetry) that emerges from the global SUSY theory is expected to be
accommodated by a string solution of the same type in the corresponding
supergravity theory, and vice versa, although the two solutions are expected
to be quantitatively different, due to the different structures of the
scalar potentials. Using these conditions, it has been shown that several
types of string solutions, such as D-type strings, F-type strings, and
superconducting Witten strings, exist in gauge invariant supergravity models.


\begin{references}
\bibitem{vsbook}  A. Vilenkin and E.P.S. Shellard, {\it Cosmic Strings and
other Topological Defects} (Cambridge University Press, 1994)

\bibitem{ktbook}  E.W. Kolb and M.S. Turner, {\it The Early Universe}
(Addison-Wesley, 1990)

\bibitem{vilenkin}  A. Vilenkin, Phys. Rep. {\bf 121}, 263 (1985)

\bibitem{cvetic1}  M. Cvetic, F. Quevedo, and S-J Rey, Phys. Rev. Lett. {\bf %
67}, 1836 (1991)

\bibitem{cvetic4}  M. Cvetic, S. Griffies, and S-J Rey, Nucl. Phys. B {\bf %
381}, 301 (1992)

\bibitem{wbbook}  See, for example, J. Wess and J. Bagger, {\it %
Supersymmetry and Supergravity}, Second Edition (Princeton University Press,
1992)

\bibitem{morriscs}  J.R. Morris, Phys. Rev. D {\bf 53}, 2078 (1996)

\bibitem{davistrod}  S.C. Davis, A.C. Davis, and M. Trodden, {\it N=1
Supersymmetric Cosmic Strings}, hep-ph/9702360

\bibitem{witten}  E. Witten, Nucl. Phys. B {\bf 249}, 557 (1985)

\bibitem{fayet2}  P. Fayet and J. Illiopoulos, Phys. Lett. B {\bf 51}, 461
(1974)

\bibitem{nielsen}  H.B. Nielsen and P. Olesen, Nucl.\ Phys. B {\bf 61}, 45
(1973)

\bibitem{vilev}  A. Vilenkin and A.E. Everett, Phys. Rev. Lett. {\bf 48},
1867 (1982)
\end{references}
\end{document}